\documentclass[reqno]{amsart}
\usepackage{graphicx}
\newcommand{\SetFigFont}[3]{}
 
\newtheorem{theorem}{Theorem}[section]

\theoremstyle{definition}

\theoremstyle{remark}

\newcommand{\spc}{\;\;\;\;\;\;\;\;\;\;}
\newcommand{\bra}{\mbox{$< \!\!$ \nolinebreak}}
\newcommand{\ket}{\mbox{\nolinebreak $>$}}
\newcommand{\beq}{\begin{equation}}
\newcommand{\eeq}{\end{equation}}

\newcommand{\Eout}{E_{\mbox{\scriptsize{\rm{out}}}}}
\newcommand{\R}{\mathbb{R}}
\newcommand{\C}{\mathbb{C}}
\newcommand{\Z}{\mathbb{Z}}

\newcommand{\1}{\mbox{\rm 1 \hspace{-1.05 em} 1}}
\numberwithin{equation}{section}

\begin{document}

\title[Linear waves in the Kerr geometry]{Linear Waves in the Kerr Geometry: \\
a Mathematical Voyage to Black Hole Physics}


\author[Finster, Kamran, Smoller and Yau]{Felix Finster, Niky Kamran, Joel Smoller and Shing-Tung Yau}
\address{NWF I -- Mathematik, Universit{{\"a}}t Regensburg, 93040 Regensburg, Germany}
\email{Felix.Finster@mathematik.uni-regensburg.de}
\thanks{Research supported in part by the Deutsche
Forschungsgemeinschaft.}

\address{Department of Mathematics and Statistics \\ McGill University \\ Montreal, QC \\ H3A 2K6 \\ Canada }
\email{nkamran@math.mcgill.ca}
\thanks{Research supported by NSERC grant RGPIN 105490-2004.}

\address{Mathematics Department \\ The University of Michigan \\ Ann Arbor, MI 48109 \\ USA}
\email{smoller@umich.edu}
\thanks{Research supported in part by the National Science Foundation, Grant No.~DMS-0603754.}

\address{Mathematics Department \\ Harvard University \\ Cambridge, MA 01238\\USA }
\email{yau@math.harvard.edu}
\thanks{Research supported in part by the NSF, Grant No.\ 33-585-7510-2-30.}



\dedicatory{}

\begin{abstract}
This paper gives a survey of wave dynamics in the Kerr space-time geometry,
the mathematical model of a rotating black hole in equilibrium.
After a brief introduction to the Kerr metric, we review the separability properties
of linear wave equations for fields of general spin~$s=0,\frac{1}{2}, 1, 2$, corresponding to scalar,
Dirac, electromagnetic fields and linearized gravitational waves.
We give results on the long-time dynamics of Dirac and scalar waves, including decay rates for
massive Dirac fields. For scalar waves, we give a rigorous treatment of superradiance
and describe rigorously a mechanism of energy extraction from a rotating black hole.
Finally, we discuss the open problem of linear stability of the Kerr metric and present partial results.
\end{abstract}

\maketitle
\tableofcontents


\section{Introduction to General Relativity and Black Holes}
It is apparently the Reverend John Michell who was the first to
suggest in a paper communicated to the Royal Society of London in
November 1783 that if one assumed that light consisted
of particles subject to the law of universal gravitation, then the
Universe could in principle contain stars that would be completely
invisible to any external observer. This conclusion was arrived at
independently by Laplace in 1796 in his {\em Exposition du syst\`eme
du monde}, and the idea is credited to him in several classical
references in General Relativity, such as {\em The Large Scale
Structure of Space-Time} by Hawking and Ellis \cite{HE}\footnote{The
reader is referred to the excellent survey by Israel \cite{I1} on
the genesis and evolution of the concept of a dark star, starting
from Michell and culminating in the laws of black hole
thermodynamics.}. The concept of a dark star would have probably
remained as a mathematical curiosity had it not been for the advent
of the Theory of Relativity and for the important advances in the
study of stellar structure which took place in the early decades of
the twentieth century, following the development of quantum
statistical mechanics. These ideas led a nineteen year old Indian
student, while on his way in July of 1930 from Madras, India, to
Cambridge, England, where he was to begin his doctoral studies, to
discover the extraordinary consequence of General Relativity
that a star which is sufficiently massive will, after having exhausted its
nuclear fuel, collapse gravitationally to a point at which the very
structure of space-time will become singular. In spite of the severe
opposition expressed to this daring work by Sir Arthur Eddington,
the leading astronomer of his time, the Indian student's work was
completely validated by further analysis and he went on to receive a
Nobel Prize many years later, in 1983, for his discovery of this
lower mass limit. His name was Subrahmanyan Chandrasekhar, and he in
turn became one of the most eminent theoretical astrophysicists of his time.

It took a long period of maturation before a proper understanding
was developed of what the space-time geometry would be near such a
collapsed object. However, it was already apparent from the early
work dating back to 1915 of Karl Schwarzschild on the relativistic
space-time geometry near a stationary spherically symmetric star,
that the singularity should not be directly visible to an external
observer who would be at rest with respect to the star, and that it
should be hidden behind a a one-way membrane, known today as an {\em
event horizon}. This led the physicist John Wheeler to refer
informally to these collapsed stars as {\em black holes}. This
evocative way of referring to these very strange objects soon became
the standard terminology.

In a lecture given in the memory of Karl Schwarzschild at the
Astronomischen Gesellschaft in Hamburg, Germany, in 1986,
Chandrasekhar stated the following \cite{CH1}:
\bigskip
\begin{quotation}
Black holes are macroscopic objects with masses varying from a few
solar masses to millions of solar masses. To the extent that they
may be considered as stationary and isolated, to that extent, they
are all, every single one of them, described {\em exactly} by the
Kerr solution. This is the only instance we have of an exact
description of a macroscopic object. Macroscopic objects, as we see
them around us, are governed by a variety of forces, derived from a
variety of approximations to a variety of physical theories. In
contrast, the only elements in the construction of black holes are
our basic concepts of space and time. They are, thus, almost by
definition, the most perfect macroscopic objects there are in the
universe. And since the general theory of relativity provides a
single two parameter family of solutions for their description, they
are the simplest as well.
\end{quotation}
\bigskip
The two parameter family of exact solutions referred to by
Chandrasekhar is the family of {\em Kerr solutions} of the Einstein
field equations of General Relativity, discovered by Roy Kerr in
1963, \cite{K}. These generalize the Schwarzschild geometry to the
case in which the black hole has non-zero angular momentum, and are
stationary and axi-symmetric as opposed to stationary and
spherically symmetric. In the introduction to an earlier paper,
written on the occasion of Einstein's centenary, Chandrasekhar wrote
about the Kerr solution~\cite{Ch0}:
\bigskip
\begin{quotation}
The special significance of Kerr's solution for astronomy derives
from the theorems of Carter (1971) and Robinson (1975) which
establish its uniqueness for an exact description of black holes
that occur in nature. But Kerr's solution has also surpassing
theoretical interest: it has many properties that have the aura of
the miraculous about them. These properties are revealed when one
considers the problem of the reflection and transmission of waves of
different sorts (electromagnetic, gravitational, neutrino, and
electron waves) by the Kerr black hole.
\end{quotation}
\bigskip
Regarding the ``miraculous properties'', he went on to say in the
conclusion of the same paper:
\bigskip
\begin{quotation}
What, may we inquire, are these properties? In many ways, the most
striking feature is the separability of all the standard equations
of mathematical physics in Kerr geometry.
\end{quotation}
\bigskip

The goal of this paper is to survey some recent developments in the
study of linear wave equations in Kerr geometry, and to thereby give
a description of some of the key physical properties of black holes
from a rigorous mathematical perspective. We remark that there are
also other very interesting and important results in mathematical relativity,
like for example black hole uniqueness theorems~\cite{Heu}, 
Hawking radiation~\cite{Wald, H2}, the formation of singularities~\cite{Christo2},
the positive mass theorem~\cite{SY, Witten} and the Penrose inequalities~\cite{hi, bray, bray2}.
But this paper has its main focus on the topics that we have worked on:
the dynamics and decay properties of various fields in the Kerr geometry.
Put differently, linear waves in the Kerr geometry will serve both as
vehicle and guide for our voyage to black hole physics. As stated above
by Chandrasekhar, the setting for our voyage is provided by the
space-time geometry corresponding to an exact solution of the
Einstein field equations of gravitation, namely the Kerr solution.
We shall see that the presence of a black hole and a space-time
singularity manifests itself in remarkable and often unexpected ways
through the behavior of waves in the space-time geometry of the
black hole's exterior. Among the unexpected aspects in the
behavior of waves, we shall explain that waves corresponding to
localized initial data for the Dirac and scalar wave equations in
Kerr geometry will always decay in any localized region of space as
$t$ tends to infinity, suggesting that the corresponding particles
either ``fall into the black hole'' or ``escape to infinity.''
Some other features are that massive Dirac waves
decay in Kerr geometry at a rate {\em slower} than in
Minkowski space, and the
fact that spatially localized wave packets can be used to extract
energy and angular momentum from the Kerr black hole.
We shall also see that even though the
geometric properties of the Kerr metric make it possible to solve all
the known linear wave equations in Kerr geometry by separation of
variables, it also gives rise to specific challenges in the analysis
of these wave equations, which require the development of new techniques.
These challenges stem from the presence of angular momentum in the
Kerr black hole, which in turn causes the conserved energy for
particles and waves to not be everywhere positive. But we are
anticipating what is to come, and before talking about black holes,
we need to set the stage by recalling some of the essentials of
General Relativity, and even before that, of Special Relativity.

Space-time in Special Relativity is given by {\em Minkowski
space-time}, that is $\mathbb{R}^{4}$, endowed with the standard
inner product of Lorentzian signature $(+,-,-,-)$. Of particular
significance are the {\em{light cones}}, consisting of the sets of
points satisfying an equation of the form
\begin{equation}\label{nullcone}
(t-t_{0})^{2}-(x-x_{0})^{2}-(y-y_{0})^{2}-(z-z_{0})^{2}=0.
\end{equation}
The light cone \eqref{nullcone} describes the set of points swept out
in space-time by a light signal emanating from the point
$(x_{0},y_{0},z_{0})$ at time $t_{0}$. (We always work in units in which
the speed of light is equal to one.) The famous Michelson-Morley
experiment showed that the speed of light is the same for all
observers in uniform motion. In particular, the observers cannot
move faster than light, and
therefore their space-time trajectories, or world lines,
lie everywhere inside the light cone. Such lines are called {\em
time-like}. The generators of the light cone \eqref{nullcone}
correspond to the world lines of light rays, and are referred to as
{\em null lines}. The light cones thus introduce a causal structure
in space-time, according to which every observer in uniform motion
has a future and a past, given by the interior of the upper,
respectively lower sheet of the light cone based at its space-time position.
The set of affine transformations of Minkowski space-time preserving
the Lorentzian inner product (and in particular the light cones)
forms the so-called Poincar\'e group, the semi-direct product of the
Lorentz group $O(1,3)$ by the group $\mathbb{R}^{4}$ of space-time
translations. All of the remarkable physical consequences of Special
Relativity, such as length contraction, time dilation and red shift
correspond to the mathematical properties of the Poincar\'e group.


While Minkowski space is a model of space-time which is very well
suited to the study of electromagnetic phenomena such as the
propagation of light, it is not sufficiently general to give a
description of space-time that includes gravitation. Indeed, the
structure of Minkowski space-time is anchored around the
privileged role played by observers in uniform motion and
therefore by linear changes of coordinates, but we already know
from Newtonian physics that gravitation manifests itself by
causing observers in free fall to be accelerated. General Relativity
emerges from Special Relativity by incorporating the Principle of
Equivalence, which equates accelerated frames of reference with
the presence of a gravitational field and allows in particular for
general local coordinate systems, which retain the Minkowskian
character of space-time. In this way, General Relativity becomes
a geometric theory of gravitation, in which space-time is taken to be a
four-dimensional manifold~$M_{4}$, endowed with a pseudo-Riemannian
metric $g$ of Lorentzian signature $(+,-,-,- )$, implying
that the tangent space at each point of space-time is
isometric to Minkowski space-time. The world lines of
test particles with non-zero rest mass acted upon only by
gravity now become {\em time-like geodesic curves} in $(M_{4},g)$,
i.e.\ geodesics~$\gamma$ whose tangent vector satisfies
\[ 0 \;<\; g(\dot \gamma,\dot \gamma) \;\equiv\; g_{ij}(\gamma)\: \dot \gamma^i\; \dot \gamma^j\:, \]
where according to Einstein's summation convention we sum over
repeated indices.
Likewise, light rays propagate along {\em null geodesics}
with~$g(\dot \gamma,\dot \gamma)=0$.
These geodesic curves are the analogue of the time-like and null
lines of Minkowski space-time.
In a local coordinate system the geodesic equation is given by
\beq \label{geodesic}
\frac{d^{2}x^{i}}{ds^{2}}+\Gamma^{i}_{{\phantom
i}kj}\frac{dx^{k}}{ds}\frac{dx^{j}}{ds}=0\:,
\eeq
where the coefficients $\Gamma^{k}_{{\phantom i}ij}$, known as the
{\em Christoffel symbols}, are built from the metric and its first
derivatives,
\beq \label{christoffel}
\Gamma^{k}_{{\phantom i}ij} \;=\; \frac{1}{2}\: g^{kr}
\left(\partial_j g_{ir}+ \partial_i g_{jr}- \partial_r g_{ij} \right).
\eeq
The geodesic equation can also be written as
\begin{equation}\label{geod}
\nabla_{\dot \gamma}{\dot \gamma} \;=\; 0\:,
\end{equation}
where~$\nabla$ is the Levi-Civita connection, often referred to as a {\em{covariant
derivative}}, defined for a general vector field~$X$ by
\[ \nabla_i X \;=\; \left( \partial_i X^j + \Gamma^j_{\;ik} \:X^k \right) \frac{\partial}
{\partial x^j}\:, \]
where~$\partial/\partial x^k$ denotes the coordinate basis of the tangent space.
The Levi-Civita connection is
the unique metric, torsion-free connection on the Lorentzian manifold.

In 1915 Albert Einstein discovered the relativistic field equations
of gravitation, which account for the presence of the gravitational field
through space-time curvature. To formulate these equations, we need
to introduce the {\em Riemann curvature tensor}, which quantifies
how a space-time $(M_{4},g)$ deviates from being Minkowskian in a
neighborhood of any point. The Riemann curvature tensor~$R^l_{\;\:ijk}$
is defined by the relations
\[ \nabla_i \nabla_j X - \nabla_j \nabla_i X
\;=\; R^l_{\;\:ijk} \:X^k\:\frac{\partial}{\partial x^l} \:, \]
valid for any vector field~$X$. It measures the degree to which the
covariant derivatives fail to commute. One can easily
see from the expression of the Christoffel symbols~\eqref{christoffel} that the
Riemann tensor is linear in the second derivatives of the components
$g_{ij}$ of the metric, and quadratic in the first derivatives. Its
expression in local coordinates is quite complicated, and we will
not give it here. What we will retain is the interpretation of the
Riemann tensor as a measure of the ``non-flatness'' of a metric, by
recalling the classical classical theorem of Riemannian geometry
which says that the condition~$R^{l}_{{\phantom k} k i j} \equiv 0$
is equivalent to $(M_{4},g)$ being locally isometric to Minkowski
space-time.

Gravitational forces can be understood in General Relativity as tidal
forces acting on nearby point particles. Mathematically, this is made
precise by the \emph{Jacobi equation}, which shows how curvature affects the behavior
of neighboring geodesics. Consider thus a one parameter family of time-like
geodesics $\gamma\,:\,(s, \alpha)\mapsto \gamma(s,\alpha)$, where
$s$ is arclength, and $\alpha$ is a parameter
labeling the geodesic curves in our family.
The vector field~$U:=\partial \gamma/\partial s$ for fixed~$\alpha$
is the unit tangent vector along the corresponding geodesic,
while the vector field~$V:=\partial \gamma/\partial \alpha$
restricted to a geodesic curve $\gamma(s,\alpha_{0})$ of our family
measures the deviation between that curve and the neighboring
geodesics. The Jacobi equation expresses the second derivative of $V$ along the geodesic
in terms of the Riemann curvature tensor,
\[ \frac{d^{2}V^{i}}{ds^{2}} \;=\; -R^{i}_{{\phantom i}kjl} \:U^{k}\,V^{j}\,U^{l}\:. \]
The right side of this equation measures the tidal force for an observer
moving along the geodesic.

Einstein discovered the field equations of gravitation by
the requirement that one recovers Newtonian gravity in the non-relativistic
limit (i.e.\ small perturbations of Minkowski space and particles
moving slowly compared to the speed of light), and taking into account the
conservation laws for energy and momentum. The field equations read
\begin{equation}\label{Einsteingrav}
R_{ij}-\frac{1}{2}\,R\,g_{ij} \;=\; 8\pi\,T_{ij}.
\end{equation}
The left-hand side of~\eqref{Einsteingrav} is
built from the {\em Ricci tensor} $R_{ij}$
and {\em scalar curvature} $R$, which are obtained by taking the trace of
the Riemann tensor,
\[ R_{ij}=R^{l}_{{\phantom k}i l j}\:,\qquad R=R^{i}_{{\phantom l}i}. \]
The right-hand side of \eqref{Einsteingrav} is the {\em energy-momentum
tensor} of the matter fields.
The left hand-side of~\eqref{Einsteingrav} is a symmetric tensor,
which is divergence-free\footnote{One
could of course add any constant multiple of the metric tensor
$g_{ij}$ to the left-hand side of \eqref{Einsteingrav} and still
obtain a tensor with zero divergence. This amounts to adding a
cosmological constant to the Einstein equations, but this constant
is usually taken to be zero in applications to black holes, where
space-time is assumed to be asymptotically flat.}
according to the second Bianchi identities, a set of integrability conditions satisfied by the
Riemann tensor. As a consequence, the energy-momentum tensor
must also be symmetric and divergence-free; this expresses the
local conservation of energy and momentum.
The Einstein field equations
need to be augmented by a set of field equations for the
matter fields. For example, in the case of the interaction of
gravity with electromagnetism, one obtains the source-free
\emph{Einstein-Maxwell equations},
\begin{gather}
R_{ij}=8\pi(F_{ik}F^{k}_{\phantom i
j}-\frac{1}{4}g_{ij}F_{kl}F^{kl}),\label{EM1}\\
\nabla_{l}F^{kl}=0,\quad \nabla_{i}F_{jk }+\nabla_{j}F_{ki
}+\nabla_{k}F_{ij}=0 \:, \label{EM2}
\end{gather}
where~$F_{ij}$ denotes the electromagnetic field tensor (and we always use the
metric to raise or lower indices).

The Einstein field equations \eqref{Einsteingrav} are a highly
complicated system of non-linear partial differential equations, which are
extremely difficult to analyze in full generality. This holds true
even in the vacuum case ($T_{ij} \equiv 0$), where they reduce to
the vacuum Einstein equations
\begin{equation}\label{Einsteinvac}
R_{ij}=0.
\end{equation}
An important analytic result on the Einstein equations was obtained
in the work by Christodoulou and Klainerman~\cite{CK},
where the nonlinear stability of Minkowski space is proved. This work has inspired
intensive research on the analysis of the Einstein equations, with and without
matter. Since in this article we cannot enter all aspects of mathematical relativity,
we shall restrict attention to special solutions of the Einstein equation and to the analysis
of {\em{linear}} wave equations in a given space-time geometry.

The first non-flat exact solution of these equations is a one-parameter family of static and
spherically symmetric metrics discovered by Karl Schwarzschild in
1915,
\begin{equation}\label{Schw}
ds^{2} \;=\; \left(1-\frac{2M}{r} \right) dt^{2}-
\left(1-\frac{2M}{r} \right)^{-1} dr^2 -r^{2}(d\theta^{2}+\sin^{2}\theta\,d\varphi^{2})\:.
\end{equation}
This metric describes a black hole of mass~$M>0$. The metric
is singular at~$r=2M$, but this singular behavior is just an artifact of the coordinate
system being used. Indeed, the metric can be extended analytically
across the event horizon by introducing the Regge-Wheeler coordinate
\[ u \;=\; r+2m\log(r-2M), \]
and an advanced null coordinate
\[ v \;=\; t+u\:, \]
in which the metric is given by
\[ ds^{2} \;=\; \left(1-\frac{2M}{r} \right)
dt^{2}-2dvdr-r^{2}(d\theta^{2}+\sin^{2}\theta\,d\varphi^{2}). \]
Now the locus~$r=2M$ is a null hypersurface which corresponds to the
boundary of the black hole and is called the {\em event horizon}.
Working in these coordinates, one can show that future-directed
time-like or null curves can only cross the hypersurface $r=2M$ from
outside to inside, which explains why this hypersurface is referred
to as an event horizon. Physically speaking, the event horizon
defines the infinite red-shift surface as seen by a distant observer.
Unlike the event horizon $r=2M$, the locus
$r=0$ is a true curvature singularity, which is not removable by any
change of coordinates. One can for example verify that the square of
the Riemann tensor blows up at~$r=0$, $R_{ijkl} R^{ijkl} \sim M^2 r^{-6}$.

According to Noether's theorem, there is a close connection between symmetries of space-time
and conservation laws. In general relativity, this connection is made precise using the notion of
a {\em{Killing field}}. A Killing vector field~$X$ is characterized by the so-called
Killing equation
\[ \nabla_i X_j + \nabla_j X_i = 0 \:. \]
This equation implies that the Lie derivative of the metric vanishes, meaning that the flow
of the vector field~$X$ is an isometry of space-time. In order to get the corresponding conservation
law, one contracts the energy momentum tensor with the Killing field.
The resulting vector field~$Y^i = T^{ij} X_j$ is divergence-free. Namely,
\[ \text{div} (Y) = \nabla_i (T^{ij} X_j) = (\nabla_i T^{ij}) \,X_j + T^{ij} \:\nabla_i X_j = 0\:, \]
where we applied the Killing equation and used 
that the energy-momentum tensor is symmetric and divergence-free as
a consequence of the Einstein equations. Applying Gauss' theorem, one concludes that
every boundary integral of~$Y$ vanishes,
\beq \label{gauss}
\int_{\partial V} Y_i \,\nu^j \:d\mu_{\partial V} = 0
\eeq
(where~$V$ is any open subset of space-time with smooth boundary, $\nu_j$ denotes a unit normal,
and~$d\mu_{\partial V}$ is the volume form on~$\partial V$).
Taking for~$V$ a set whose boundary consists of two space-like hypersurfaces corresponding
to two time slices of an observer, \eqref{gauss} gives rise to a {\em{conservation law}}
for a spatial integral.

We remark that the above method is also useful in the situation when space-time has no obvious
symmetries. In this case, one can try to find a vector field which is an approximate Killing
field in the sense that the expression~$\nabla_i X_j + \nabla_j X_i$ vanishes up to a small
error term. Then in~\eqref{gauss} one also gets error terms, but nevertheless one can hope
to obtain useful estimates. This is the so-called {\em{vector field method}} which has
been used in many papers to study the behavior of waves in the Schwarzschild geometry
(see for example~\cite{Blue, DR3, DR4}). An extension of the method to conformal symmetries
goes back to Morawetz~\cite{morawetz}.
For a general exposition on the role of vector fields in the analysis of Euler-Lagrange systems
of partial differential equations of hyperbolic type we refer to~\cite{Christo3}.

\section{The Kerr Metric and the Black Hole Uniqueness Theorem}
It took nearly fifty years until an exact solution of the Einstein
equations describing the outer geometry of a {\em rotating} black
hole in equilibrium was found. This is the Kerr metric given by
\begin{equation}\label{KN}
ds^{2} \;=\; \frac{\Delta}{U} \,( dt-a\sin^{2}\vartheta
\,d\varphi)^{2}-U \left( \frac{dr^{2}}{\Delta}+d\vartheta^{2}
\right) -\frac{\sin^{2}\vartheta}{U} \Big( a\,dt-(r^{2}+a^{2})d\varphi
\Big)^{2},
\end{equation}
where
\begin{equation}
U \;=\; r^{2}+a^{2}\cos^{2}\vartheta,\qquad \Delta \;=\; r^{2}-2Mr+a^{2},
\end{equation}
and the coordinates~$(t,r, \vartheta, \varphi)$ are in the range
\begin{equation}
-\infty<t<\infty,\;\;\;
M+{\sqrt{M^{2}-a^{2}}}<r<\infty,\;\;\; 0<\vartheta<\pi,\;\;\;
0<\varphi<2\pi.
\end{equation}
Here the parameters~$M$ and~$a$ describe the mass and the angular
momentum per unit mass of the black hole. It is easily verified that in the
case~$a=0$, one recovers the Schwarzschild metric \eqref{Schw}.

For this metric to describe a black hole we need to assume
that~$M^{2}> a^{2}$, giving a bound for the angular momentum
relative to the mass. In this so-called {\em{non-extreme case}},
it can be verified by an argument similar to the one used for the
Schwarzschild solution that the null hypersurface
\begin{equation}
r \;=\; r_{1} \;:=\; M+{\sqrt{M^{2}-a^{2}}}
\end{equation}
defines the event horizon, the boundary of the black hole.
Here we shall only consider the region~$r > r_1$ outside the event horizon.
The coefficients of the Kerr metric are independent of~$t$ and~$\varphi$,
showing that the space-time geometry is stationary and axi-symmetric.
One of the key features of the Kerr geometry is the existence of an {\em
ergosphere}, that is a region which lies outside the event horizon,
in which the vector field $\frac{\partial}{\partial t}$
is {\em space-like}. In order to determine the ergosphere, we consider the norm of the
vector field $\xi=\frac{\partial}{\partial t}$,
\begin{equation}
g_{ij}\,\xi^{i}\xi^{j} \;=\; g_{tt} \;=\;
\frac{\Delta-a^{2}\,\sin^{2}\vartheta}{U} \;=\; \frac{r^2 - 2Mr +
a^2 \cos^2 \vartheta}{U}\: .
\end{equation}
This shows that $\xi$ is space-like in the open region of space-time
where
\begin{equation}
r^{2}-2Mr+a^{2}\,\cos^{2}\vartheta \;<\; 0 \:,
\end{equation}
which defines the ergosphere. It is a bounded region of space-time
outside the event horizon, which intersects the event horizon at the
poles $\vartheta=0,\,\pi$.
The implications of the ergosphere for the
analysis of wave equations cannot be overstated because the ergosphere
corresponds to a region in which the conserved energy associated to
a given field will {\em fail to be positive} (see for example~\eqref{energy}
and~\eqref{energydensity} below).

The physical significance of the Kerr metric manifests itself through
the {\it black hole uniqueness theorems} of Israel \cite{I}, Carter \cite{C0}
and Robinson~\cite{R, Heu}. Physically, these theorems indicate that the
stationary end-state of a rotating black hole should be given precisely by
the non-extreme Kerr geometry.
From the mathematical point of view, these theorems are uniqueness
results for a class of boundary value problems for the vacuum Einstein
equations, and their proof requires additional mathematical assumptions.
As boundary conditions one assumes the asymptotic flatness of space-time in the
form of weak asymptotic simplicity (see~\cite[page~282]{Wald}),
plus the existence of an event horizon with the natural spherical topology.
Furthermore, one assumes axi-symmetry and pseudo-stationarity
(pseudo-stationarity means that there is a Killing field which is assumed to be
 timelike only in the asymptotic end).
Finally, as technical assumptions which are also motivated from physics, one needs to
impose a causality condition as well as time orientability, and one needs to
fix the topology of space-time to be~${\mathbb{R}}^2 \times S^2$.
(The review article by Carter~\cite{C00} and the books
\cite{HE, Heu} specify and discuss all the conditions in detail.)
The theorem is as follows:
\begin{theorem}
Under the above assumptions, every solution of the vacuum Einstein equations
admits a global chart $(t,r,\vartheta,\varphi)$ in
which the metric is the Kerr solution~\eqref{KN}.
\end{theorem}

\section{Linear Wave Equations in the Kerr Geometry and their Separation}\label{Separ}
In this section we give an overview of the physically relevant
linear wave equations and present their separability properties in the Kerr geometry.
For simplicity, we begin with the scalar wave equation,
which in a general space-time reads
\begin{equation} \label{swave}
g^{ij}\nabla_{i}\nabla_{j}\,\Phi=\frac{1}{\sqrt
{-g}}\,\frac{\partial}{\partial x^{i}} \left(
{\sqrt{-g}}\,g^{ij}\frac{\partial}{\partial x^{j}} \right) \Phi
\;=\; 0 \:,
\end{equation}
where $g$ denotes the determinant of the metric $g_{ij}$. For the
Kerr metric, this becomes
\begin{eqnarray} \label{wave}
\lefteqn{ \left[ -\frac{\partial}{\partial
r}\Delta\frac{\partial}{\partial r} +\frac{1}{\Delta} \left(
(r^{2}+a^{2})\frac{\partial}{\partial t}+a\frac{\partial}{\partial
\varphi} \right)^{2}  \right. } \nonumber \\
&& \left. -\frac{\partial}{\partial \cos \vartheta}
\sin^{2}\vartheta
\frac{\partial}{\partial \cos \vartheta}
-\frac{1}{\sin^{2}\vartheta} \left(
a\sin^{2}\vartheta \frac{\partial}{\partial
t}+\frac{\partial}{\partial \varphi}\right)^{2} \right ]\Phi \;=\;
0\:.
\end{eqnarray}
We denote the square bracket in this equation by~$\square$,
(although it is actually a scalar function times the wave operator
in~\eqref{swave}). We now explain the separation of variables
as discovered by Carter~\cite{C00}.
Due to the stationarity and axi-symmetry, we can separate the
$t$- and $\varphi$-dependence with the usual plane-wave ansatz
\begin{equation}\label{separansatz}
\Phi(t,r,\vartheta,\varphi) \;=\; e^{-i\omega
t-ik\varphi}\:\phi(r,\vartheta)\:,
\end{equation}
where $\omega$ is a quantum number which could be real or complex
and which corresponds to the ``energy'', and $k \in {\mathbb{Z}}$ is
a quantum number corresponding to the projection of angular momentum
onto the axis of symmetry of the black hole.
Substituting~\eqref{separansatz} into~\eqref{wave}, we see that the
wave operator splits into the sum of radial and angular parts,
\begin{equation} \label{wavesep}
\square \Phi \;=\; ({\mathcal{R}}_{\omega , k}+{\mathcal{A}}_{\omega
, k}) \Phi,
\end{equation}
where ${\mathcal{R}}_{\omega , k}$ and ${\mathcal{A}}_{\omega , k}$
are given by
\begin{eqnarray} \label{radial}
{\mathcal{R}}_{\omega , k} &=& \, -\frac{\partial}{\partial r}
\Delta \frac{\partial}{\partial r}
- \frac{1}{\Delta}((r^{2}+a^{2})\omega +ak)^{2}, \\
{\mathcal{A}}_{\omega , k} &=& \label{angular}
-\frac{\partial}{\partial \cos \vartheta}\: \sin^2 \vartheta\:
\frac{\partial}{\partial \cos \vartheta} +\frac{1}{\sin^2
\vartheta}(a\omega \sin^2 \vartheta + k )^{2} \:.
\end{eqnarray}
We can therefore separate the variables $r$ and $\vartheta$
with the multiplicative ansatz
\beq \label{lastsep}
\phi(r, \vartheta) \;=\; R(r)\: \Theta(\vartheta)\:,
\eeq
to obtain for given $\omega$ and $k$ the system of ordinary
differential equations
\begin{equation} \label{odes}
{\mathcal{R}}_{\omega ,k}\,R_\lambda \;=\; -\lambda \,
R_{\lambda},\qquad  {\mathcal{A}}_{\omega ,k}\,\Theta_\lambda \;=\;
\lambda \, \Theta_\lambda\:.
\end{equation}
The separation constant $\lambda$ is an eigenvalue of the angular
operator ${\mathcal{A}}_{\omega, k}$ and can thus be thought of as
an angular quantum number. In the spherically symmetric
case (i.e.\ $a=0$), $\lambda$ goes over to the usual eigenvalues
$\lambda=l(l+1)$ of the total angular momentum.
We point out that the last separation~\eqref{lastsep} is not obvious
because it does not correspond to an underlying space-time symmetry.

For the Dirac equation describing a spin $\frac{1}{2}$ field,
the situation is more complicated because
it is a system of differential equations and involves a rest mass~$m$.
The separability of the Dirac equation in the Kerr metric was
first established by Chandrasekhar in 1976,~\cite{Ch}, using an ingenious
new method. The Dirac equation for a particle of mass~$m$ reads
\begin{equation}\label{Dirac}
(i \gamma^{j}\nabla_{j}-m)\Psi \;=\; 0,
\end{equation}
where the Dirac matrices~$\gamma^j$ are related to the metric by the anti-commutation
relations
\[ \frac{1}{2} \left( \gamma^j \gamma^k + \gamma^k \gamma^j \right) \;=\; g^{jk} \:\1\:, \]
and~$\nabla$ is a connection on the spinors.
We write the Dirac equation in a Newman-Penrose null frame~\cite{Ch} $(l,n,m,{\bar m})$,
i.e.\ in a frame in which the metric takes the form
\[ g_{ij}=l_{i}n_{j}+n_{i}l_{j}-m_{i}{\bar m}_{j}-{\bar m}_{i}m_{j} \:. \]
Choosing a suitable frame and after an appropriate linear transformation of the Dirac spinor,
\[ \Psi \;\to\; \hat{\Psi} \;=\; S\: \Psi \:, \]
where~$S=S(r,\vartheta)$ is a diagonal matrix, the Dirac equation becomes
\begin{equation}
({\mathcal{R}}+{\mathcal{A}}) \:\hat{\Psi} \;=\; 0 \:, \label{eq:1}
\end{equation}
where $\mathcal{R}$ and $\mathcal{A}$ are certain first-order
matrix differential operators. The separation of variables is
achieved by assuming that each component of the transformed spinor
$\hat{\Psi}$ is a product of functions of one variable, according
to a specific separation pattern,
\begin{equation}
\hat{\Psi}(t,r,\vartheta,\varphi) \;=\; e^{-i \omega t} \:e^{-i
(k+\frac{1}{2}) \varphi}
\: \left( \begin{array}{c} X_-(r) \:Y_-(\vartheta) \\
X_+(r) \:Y_+(\vartheta) \\
X_+(r) \:Y_-(\vartheta) \\
X_-(r) \:Y_+(\vartheta) \end{array} \right) \;,\spc k\in \mathbb{Z}.
\label{eq:21}
\end{equation}
By substituting \eqref{eq:21} into \eqref{eq:1}, we obtain the
separated matrix equations
\begin{equation}
{\mathcal{R}} \:\hat{\Psi} \;=\; \lambda\:\hat{\Psi} \;,\spc
{\mathcal{A}} \:\hat{\Psi} \;=\; -\lambda\:\hat{\Psi} \;,
\label{eq:23a}
\end{equation}
where $\lambda$ is a separation constant, under which the
transformed Dirac equation \eqref{eq:1} decouples into a system of
ODEs.

The remaining field equations of physical interest are those describing
electromagnetic and linearized gravitational waves. It is a
remarkable and very useful fact that these waves are all governed by a single
second-order equation, the so-called \emph{Teukolsky master
equation} \cite{Ch}. It reads
\begin{equation} \label{Teukolsky}
\begin{split}
\lefteqn{ \left[ \frac{\partial}{\partial r}\Delta\frac{\partial}{\partial r}
-\frac{1}{\Delta} \left( (r^{2}+a^{2})\frac{\partial}{\partial
t}+a\frac{\partial}{\partial \phi}-(r-M)s \right)^{2} -4s(r+ia\cos
\theta)\frac{\partial}{\partial t} \right. } \\
&\;\;\left. +\frac{\partial}{\partial \cos \theta} \sin^{2}\theta
\frac{\partial}{\partial \cos
\theta}+\frac{1}{\sin^{2}\theta} \left( a\sin^{2}\theta
\frac{\partial}{\partial t}+\frac{\partial}{\partial \phi}+is\cos
\theta \right)^{2} \right]\Psi_{s} \;=\; 0\:.
\end{split}
\end{equation}
The parameter~$s$ denotes the spin.
We explain how this equation comes about, beginning
with the case~$s=\pm 1$ of a Maxwell field.  The source-free
Maxwell's equations are given by
\begin{equation}\label{Maxwell}
\nabla_{l}F^{kl}=0,\qquad \nabla_{i}F_{jk }+\nabla_{j}F_{ki
}+\nabla_{k}F_{ij}=0\:,
\end{equation}
where~$F_{ij}$ is the electromagnetic field tensor.
Choosing the same Newman-Penrose null frame as in the separation of
the Dirac equation, we can combine the components of the field tensor
into the three complex functions
\[ \Psi_{1} \;=\; F_{ij}l^{i}m^{j},\quad
\Xi_0 \;=\; \frac{1}{2 \rho}\: F_{ij} \left( l^i n^j + {\bar m}^i m^j \right) ,\quad
\Psi_{-1} \;=\; \frac{1}{\rho^2}\: F_{ij}{\bar m}^{i}n^{j}, \]
where~$\rho=-(r-ia\cos\vartheta)^{-1}$. Then
Maxwell's equations~\eqref{Maxwell} can be written as
\[ {\mathcal{D}} \left(\!\! \begin{array}{c} \Psi_1 \\ \Xi_0 \\ \Psi_{-1}
\end{array} \!\!\right) \;=\; 0 \:, \]
where~${\mathcal{D}}$ is a first-order matrix differential operator.
The key point is that, multiplying this equation from the left by a suitable
first-order matrix differential operator, one obtains decoupled second order
equations for~$\Psi_1$ and~$\Psi_{-1}$, which are precisely the Teukolsky
equation for~$s=\pm 1$.
It is important to note that this process of cross-differentiation
and elimination makes use of some key commutation identities
between certain covariant derivative operators in the Kerr metric.
These identities hold as a consequence of the special algebraic
structure of the Riemann tensor in the Kerr geometry; namely,
it is of type D in the Petrov-Penrose classification,
meaning that the Weyl tensor has two repeated principal null
directions. If~$\Psi_1$ or~$\Psi_{-1}$ are known, then the
remaining components are readily obtained by differentiation
using the so-called Teukolsky-Starobinsky identities~\cite{Ch}.

Finally, for $s=\pm 2$, the Teukolsky equation~\eqref{Teukolsky}
is obtained similarly by cross-differentiation and elimination
on the systems of first-order equations obtained by linearizing the
Bianchi identities
\[ \nabla_{i} R_{jklm}+\nabla_{j} R_{kilm}+\nabla_{k}R_{ijlm}=0, \]
around the Kerr metric. If we denote by $\psi_{0}^{(1)}$ and
$\psi_{4}^{(1)}$ the first-order perturbations of the
Newman-Penrose curvature components $\psi_{0}$ and $\psi_{4}$
defined by
\[
\psi_{0} \;=\; C_{ijkl}l^{i}m^{j}l^{k}m^{l},\quad
\psi_{4} \;=\; C_{ijkl}n^{i}{\bar m}^{j}n^{k}{\bar m}^{l},
\]
where $C_{ijkl}$ denotes the Weyl conformal curvature tensor, then
the unknowns in the Teukolsky equations with $s=\pm 2$ are given
by
\[
\Psi_{2}=\psi_{0}^{(1)},\quad \Psi_{-2}=\rho^{-4}\psi_{4}^{(1)}\:.
\]

We remark that for $s=0$, the Teukolsky equation~\eqref{Teukolsky}
reduces to the scalar wave equation \eqref{swave}.
The case~$s=1/2$ corresponds to the massless Dirac equation.
We also point out that the reduction to a second order scalar
equation does not work for the massive Dirac equation.
Just like the scalar wave equation, the Teukolsky
equation can be reduced to ordinary differential equations by the
separation ansatz
\[ \Psi_{s}=e^{-i\omega t-ik\varphi}\:R_{s}(r)\:\Theta_{s}(\vartheta)\:. \]
These are the separability properties that were referred to by
Chandrasekhar \cite{Ch0} in the excerpt quoted in the Introduction
as ``having the aura of the miraculous.''

Clearly, the above linear wave equations can also be analyzed in the special case~$a=0$,
where the Kerr geometry simplifies to the spherically symmetric and static
Schwarzschild geometry~\eqref{Schw}. Since in this paper we shall concentrate on
rotating black holes, we merely mention a few important papers on the analysis of
waves in the Schwarzschild geometry. In the fundametal paper~\cite{KW} it is proven that
the solutions of the scalar wave equation are pointwise bounded, uniformly in time.
The corresponding rates of decay are derived in~\cite{DR2} and~\cite{K1,K2}.
In~\cite{Blue} the boundedness of solutions to Maxwell's equations in the Schwarzschild
geometry is proved. Finally, \cite{DR1} considers the scalar wave equations in the Kerr metric in the
case~$a \ll M$ of small angular momentum, where the metric is close to Schwarzschild.

\section{Dynamics of Dirac Waves}
To get a more detailed picture of the properties of a rotating black hole, it is
very helpful to consider dynamical situations where physical objects are moving
in the Kerr geometry. The simplest situation is to consider point particles,
whose motion is described by the geodesic equation~\eqref{geodesic}.
As is worked out in detail in~\cite{Ch}, there are stable orbits where a
point particle ``rotates around'' the black hole. Apart from the circular
orbits, there are elliptic orbits which are not quite closed due to the
perihelion shift. Furthermore, there are orbits where the particle falls into
the black hole or escapes to infinity.
Considering quantum mechanical particles makes the situation more interesting,
because due to the Heisenberg Uncertainty Principle, these phenomena could
happen simultaneously with certain probabilities. To clarify the picture,
in this section we consider the dynamics of Dirac waves in the Kerr geometry.
We shall first see how the Cauchy problem for
the Dirac equation in Kerr geometry can be solved by means of an
integral representation for the propagator, realized as a
superposition of ``modes'' arising from the separation of
variables. We then analyze this integral representation to show
that in the presence of a black hole, a quantum mechanical Dirac
particle cannot remain in a localized region of space for
arbitrarily large times, in other words that the quantum
mechanical particle corresponding to the Dirac wave function will
either eventually ``fall into the black hole'' or ``escape to
infinity''. We also discuss the decay rates.
Our analysis is of course
limited by the fact that we look at a test Dirac field in Kerr
geometry as opposed the fully coupled axisymmetric Einstein-Dirac
equations, but the non-existence results proved in \cite{FSY} for
black hole solutions of the spherically symmetric
Einstein-Dirac-Maxwell equations suggest that the same conclusions
should hold in the fully coupled axisymmetric
case\footnote{A major obstacle in the analysis of the fully
coupled axisymmetric Einstein-Dirac-Maxwell equations by similar
methods is that a complete separation of variables into ordinary
differential equations seems most unlikely, and one has to deal
with the full system of PDEs.}.

The first step is to bring the Dirac equation \eqref{eq:1} into
Hamiltonian form, in a way that is compatible with the separation
of variables. This is not a trivial step because the coefficients
of the $\partial_t$-term depend on both~$r$ and~$\vartheta$.
The Dirac equation becomes
\beq \label{schro}
i \:\frac{\partial}{\partial t} \:\Psi \;=\; H \: \Psi \:,
\eeq
where~$H$ is a first-order matrix differential operator.
It is important for the analysis of this evolution equation
to introduce a positive scalar product $\bra .|. \ket$,
with respect to which~$H$ is symmetric.
To this end, we consider the Dirac current~${\overline \Psi} \gamma^{j}\Psi$,
which, as a consequence of the Dirac equation, it is divergence free.
Furthermore, it is timelike and future-directed, so that
its inner product $\overline{\Psi} \gamma^0 \Psi$ with the unit normal
to the space-like hypersurface~$t={\mbox{const}}$ is positive,
even inside the ergosphere.
We introduce~$\bra \Psi \:|\Phi \ket$ by integrating the
corresponding bilinear form $\overline{\Psi} \gamma^0 \Phi$
over the hypersurface~$t={\mbox{const}}$. Then the Gauss
divergence theorem yields that~$\bra \Psi \:|\: \Phi \ket$
is time independent. Hence, using~\eqref{schro},
\[ 0 \;=\; \frac{d}{dt} \bra \Psi(t) \:|\: \Phi(t) \ket
\;=\; i \left( \bra H \Psi(t) \:|\: \Phi(t) \ket
- \bra \Psi(t) \:|\: H \Phi(t) \ket \right) , \]
and thus~$H$ is indeed a symmetric operator on the
corresponding Hilbert space~${\mathcal{H}}$.

After constructing a self-adjoint extension, we can use the
functional calculus and the spectral theorem in Hilbert spaces to obtain
\[ \Psi(t) \;=\; e^{-i t H} \Psi_0 \;=\;
\int_{\sigma(H)} e^{-i \omega t}\: dE_\omega\:. \]
Using the separation of variables, we can analyze the spectrum
and the spectral measure.
More precisely, as a technical tool for analyzing the spectral measure,
we first set up the problem in a finite radial box, with Dirichlet boundary conditions
on the walls of the box that make the Dirac operator essentially
self-adjoint. The Hamiltonian for this problem has a purely
discrete spectrum and one thus obtains a spectral representation
of the propagator as an infinite sum of discrete eigenstates,
expressible as products of eigenfunctions of the radial and angular
operators arising from the separation of variables.
The next step is to take limits as the walls of the box tend to the event
horizon and to infinity. In these limits, the sum over the discrete eigenvalues goes over to
an integral representation for the solution of the Cauchy problem.
We thus obtain the following integral representation.
\begin{theorem}\label{intrepDirac}
Consider the Cauchy problem for the Dirac equation in the
Kerr geometry,
\begin{equation}
(i \gamma^{j}\nabla_{j}-m) \Psi(t,x) \;=\; 0\:,\qquad
\Psi(0,x)=\Psi_{0}(x), \label{dcauchy}
\end{equation}
for initial data~$\Psi_0 \in C^{\infty}_{0}((r_1, \infty) \times S^{2})^{4}$.
Then
\begin{equation}\label{Fourier}
\Psi(t,x) \;=\; \frac{1}{\pi}\sum_{k,n\in
\mathbb{Z}}\int_{-\infty}^{+\infty}e^{-i\omega
t}\sum_{a,b=1}^{2}t^{k\omega n}_{ab}\Psi^{k\omega n}_{a}(x)\bra
\Psi^{k\omega n}_{b}\:|\:\Psi_{0}\ket\, d\omega,
\end{equation}
where the sums and the integrals converge in the ${\mathcal{H}}$-norm.
\end{theorem}
\noindent
In the above integral representation, the integer $k$ is the
quantum number corresponding to the projection of the angular
momentum on the axis of symmetry of the black hole, and $n$ is the
generalized total angular momentum quantum number arising from the
separation of variables, corresponding to the discrete spectrum of
the angular part of the Dirac operator in the Kerr geometry.
The $\Psi^{k\omega n}_{a}$ are solutions of the Dirac equation
which arise from the separation of variables and satisfy asymptotic
boundary conditions at infinity and near the event horizon.
More precisely, for $|\omega|>m$ the~$\Psi^k_{k \omega n}$
behave near the event horizon like incoming spherical waves for $a=1$ and outgoing
spherical waves for $a=2$, whereas in the case $|\omega| < m$,
the $\Psi^{k\omega n}_{a}$ is a linear combination of both incoming and outgoing
spherical waves near the event horizon, such that the solution has
exponential decay at infinity. The coefficients
$t^{k\omega n}_{ab}$ are given explicitly in terms of the
reflection and transmission coefficients.

Applying the Riemann-Lebesgue lemma to the above integral representation
gives the following local decay result.
\begin{theorem}\label{decay}
Consider the Cauchy problem~\eqref{dcauchy} for the Dirac equation in the
Kerr geometry,
where the initial data $\Psi_{0}$ is in $L^{2}((r_{1},\infty)\times
S^{2})^{4}$ and in $L^{\infty}_{loc}$ near the horizon, i.e.
$|\Psi_{0}(x)|< c$ for $x \in (r_{1},r_{1}+\epsilon)\times S^{2}$.
Let $\delta
> 0$ be given, let $R>r_{1}+\delta$ and consider the compact
space-like hyper-surface $K_{\delta,R}$ of $\mathcal E$ given by
\begin{equation}
K_{\delta,R}:=\{(t,r,\theta,\phi)\,|\, r_{1}+\delta\leq r \leq
R\,,t=const.\}.
\end{equation}
Then the probability for the Dirac particle to be inside
$K_{\delta,R}$ tends to zero as $t \to \infty$, that is
\begin{equation}
\lim_{t\to \infty}\int_{K_{\delta,R}}({\overline
\Psi}\gamma^0 \Psi)(t,x)\,d\mu =0,
\end{equation}
where~$d\mu$ denotes the induced volume element on $K_{\delta,R}$.
\end{theorem}

Theorem \ref{decay} implies that the Dirac spinor $\Psi$ decays to
zero in $L^{\infty}_{loc}$, or equivalently that the Dirac
particle must eventually either disappear into the black hole, or
escape to infinity. In order to get a more detailed physical picture,
one would like to determine the
probability of these outcomes in terms of the
Cauchy data. Likewise, one would like to determine the rates of
decay of the Dirac spinor in $L^{\infty}_{loc}$, as $t$ tends to
infinity. Both of these questions were addressed in~\cite{FKSY02},
under the additional assumption that in
the integral representation~\eqref{Fourier} only a finite number
of angular momentum modes are present,
\begin{equation}\label{finiteangmom}
|k|\leq k_{0}\:,\qquad |n| \leq n_{0}\:.
\end{equation}
We now state the main results, beginning with the decay rates.
\begin{theorem}\label{decayrates}
Consider the Cauchy problem as in Theorem \ref{decay}, with
initial data normalized by $\bra\Psi_{0}\:|\:\Psi_{0}\ket=1$.
Suppose that \eqref{finiteangmom} holds. Then we have:
\medskip
\item{i)} If for any $k$ and $n$,
\begin{equation}
\limsup_{\omega \searrow m}|\bra \Psi_{2}^{k\omega
n}\:|\:\Psi_{0}\ket|\neq 0,\quad  or \quad \liminf_{\omega
\nearrow -m}|\bra \Psi_{2}^{k\omega n}\:|\:\Psi_{0}\ket|\neq 0,
\end{equation}
then, as $t\to \infty$,
\begin{equation}
|\Psi(x,t)|=ct^{-5/6}+O(t^{-5/6-\epsilon}),
\end{equation}
where $c=c(x)\neq 0$, and $\epsilon < {1/{30}}$.
\medskip
\item{ii)}If for all $k,n$ and $a=1,2$, $\bra \Psi_{a}^{k\omega
n}\:|\:\Psi_{0}\ket=0$ for all $\omega$ is a neighborhood of $\pm
m$, then for any fixed $x$, $\Psi(x,t)$ decays rapidly in $t$.
\end{theorem}
\noindent
Note that the decay rate obtained in Theorem
\ref{decayrates} is \emph{slower} than the rate of decay of
$t^{-3/2}$ one obtains for the solutions of the Dirac equation in
Minkowski space, ~\cite{FKSY02}. At first glance, this result
seems to be somewhat counter-intuitive. Indeed, because the Kerr
metric is asymptotically flat, the Dirac spinor should behave
near infinity like a solution of the Dirac equation in Minkowski
space, where the rate of decay is in $t^{-3/2}$. On the other
hand, near the event horizon, the Dirac particle should behave
like a massless particle, and its wave function should have rapid
decay. One would therefore expect the rate of decay in the
Kerr metric to be \emph{at least} as fast as $t^{-3/2}$.
This naive picture is in fact incorrect. Indeed, as shown in
\cite{FKSY02}, the slower rate of decay can be understood by the
fact that in the Kerr geometry, the energy spectrum of the
initial data for the Dirac equation oscillates more and more as
$\omega$ approaches the rest mass $m$ of the Dirac particle
(essentially as $\sin((m-\omega)^{-1/2})$ as opposed to
$(\omega-m)^{1/2}$ in Minkowski space). Upon taking the Fourier
transform in time, these oscillatory contributions
lead to a slower decay rate. Another way of understanding this
effect is that the backscattering of the outgoing wave near
infinity slows down the decay.

We now turn to the probability estimates. The
probability for the particle to escape to infinity is
given by
\begin{equation}
p \;=\; \lim_{t \to \infty}
\int_{r>R}{\overline{\Psi}}\gamma^{j}\Psi(t,x)\:\nu_{j}\,d\mu,
\end{equation}
where $R>r_{1}$. In~\cite{FKSY02} it is shown that the probability $p$ is
independent of~$R$ and can be expressed in terms of the quantities appearing in the
integral representation \eqref{Fourier} by
\[ p \;=\; \frac{1}{\pi}\sum_{|k|\leq k_{0}}\sum_{|l|\leq
l_{0}}\int_{{\mathbb{R} [-m,m]}} \left(
\frac{1}{2}-2\,|t^{k\omega n}_{12}|^{2} \right)
|\bra \Psi^{k\omega n}_{2} \:|\: \Psi_{0} \ket |^{2}d\omega\:. \]
Analyzing this expression one gets the following result.
\begin{theorem}\label{probest}
Consider the Cauchy problem as in Theorem \ref{decay}, with
initial data normalized by $\bra\Psi_{0}\:|\:\Psi_{0}\ket=1$. Then
the following hold:
\begin{itemize}
\item[(i)] If the outgoing initial energy distribution satisfies
$\bra \Psi_{2}^{k\omega n}\:|\:\Psi_{0}\ket\neq 0$ for some $\omega$
such that $|\omega|>m$, then $p>0$.
\item[(ii)] If the initial energy distribution satisfies for $a=1$
or $a=2$, $\bra \Psi_{a}^{k\omega n}\:|\:\Psi_{0}\ket \neq 0$ for
some $\omega$ such that $|\omega|>m$, then $p<1$.
\item[(iii)] If the
initial energy distribution is supported in the interval $[-m,m]$,
then $p=0$.
\item[(iv)] If for any $k$ and $n$,
\begin{equation}
\limsup_{\omega \searrow m}|\bra \Psi_{2}^{k\omega
n}\:|\:\Psi_{0}\ket|\neq 0
\quad {\mbox{or}} \quad
\liminf_{\omega \nearrow -m}|\bra \Psi_{2}^{k\omega
n}\:|\:\Psi_{0}\ket|\neq 0,
\end{equation}
then $0<p<1$.
\item[(v)] We have $p=1$ if
and only if for all $k$ and $n$ the following conditions hold:
\begin{gather}
\bra \Psi_{1}^{k\omega n}\:|\:\Psi_{0}\ket =0, \quad \mbox{if}\quad
|\omega|\leq m,\\
\bra \Psi_{1}^{k\omega n}\:|\:\Psi_{0}\ket =-2t^{k\omega n}_{12}\bra
\Psi_{2}^{k\omega n}\:|\:\Psi_{0}\ket \quad \mbox{if}\quad |\omega|>
m.
\end{gather}
\end{itemize}
\end{theorem}

We point out that in the above theorems we always considered
solutions supported outside the event horizon. One can also
analyze the Dirac equation across the event horizon in the weak
sense, and show that there are no time-periodic weak solutions
which are in $L^{2}$ away from the event horizon~\cite{FSY00,
FKSY00}.

We finally mention a few other rigorous results on the dynamics of quantum
mechanical waves in the Kerr geometry which we we cannot describe in this article.
First, the scattering theory for Dirac particle has been developed in~\cite{HN, Daude, Batic}.
 A remarkable result on the Klein-Gordon equation
was obtained in~\cite{H1}.
The paper~\cite{H2} is devoted to the Hawking effect, which leads to the creation
of fermions by a rotating black hole.

\section{An Integral Representation and Decay for Scalar Waves}\label{scalardyn}
For the scalar wave equation, we derive an integral representation which is
similar to that obtained for the Dirac operator in Theorem~{\ref{intrepDirac}},
but the proof is significantly more difficult and requires new techniques.
The integral representation will imply decay as for the Dirac equation,
but it will in addition lead to the phenomenon of superradiance, which will be
explained in Section~\ref{superradiance}. In contrast to the Dirac
operator, the scalar wave operator in Kerr geometry does {\em not}
admit a conserved quantity which is positive everywhere outside
the event horizon, making it impossible to apply the spectral
theory of self-adjoint operators in Hilbert space. First of all, the
energy~$E[\Phi]$ is not positive. Namely, using the invariance of
the scalar wave Lagrangian
\beq \label{Lagrange}
{\mathcal{L}}[\Phi] \;=\; |\nabla \Phi|^2
\eeq
under time translations, we obtain by Noether's theorem the expression
\begin{equation}\label{energy}
E[\Phi] \;=\; \int_{r_1}^{\infty} dr \int_{-1}^{1} d(\cos\vartheta)
\int_{0}^{2\pi} \frac{d\varphi}{2 \pi} \:\mathcal{E},
\end{equation}
where ${\mathcal{E}}$ is the energy density
\begin{eqnarray}
{\mathcal{E}} &=& \left({\frac{(r^{2}+a^{2})^{2}}{\Delta}} -
a^{2}\,\sin^{2}\vartheta \right) \left|\partial_{t} \Phi
\right|^{2}+\Delta
\left|\partial_{r}\Phi \right|^{2} \nonumber \\
&&+\sin^{2}\vartheta \left|\partial_{\cos \vartheta}\Phi
\right|^{2}+\left(
{\frac{1}{\sin^{2}\vartheta}}-{\frac{a^{2}}{\Delta}}\right) \left|
\partial_{\varphi}\Phi \right|^{2} . \label{energydensity}
\end{eqnarray}
One sees that all the terms in the energy density are positive,
except for the coefficient of $|\partial_{\varphi}\Phi|^{2}$,
which is positive if and only if
$r^{2}-2Mr+a^{2}\,\cos^{2}\vartheta >0$, that is, precisely
outside the ergosphere. The difficulty is that the conservation of~$E[\Phi]$
does not give a Sobolev estimate for~$\Phi$. In particular, 
energy conservation does not rule out the situation where~$\Phi$ blows up in time,
in such a way that the energy density tends to minus infinity inside the ergosphere
and to plus infinity outside the ergosphere.
Moreover, in contrast to the Dirac equation, also the conserved charge is non-positive.
In~\cite{FKSY05} it is shown that no other first order or higher order positive conserved energy
exists for the scalar wave equation in the Kerr geometry.

We begin by stating a theorem which gives an integral spectral representation
for the solution of the Cauchy problem for the scalar wave
equation. This theorem can be thought of as the analogue of
Theorem~\ref{intrepDirac} for the scalar wave equation.
\begin{theorem}\label{intrepscal}
Given initial data $\Psi_0 \in C^\infty_0(\mathbb{R} \times
S^2)^2$, the solution of the Cauchy problem for the scalar wave
equation \eqref{wave} can be represented as
\begin{equation}\label{intrep} \Psi(t,r,\vartheta,
\varphi) \;=\; \frac{1}{2 \pi} \sum_{k \in \mathbb{Z}} e^{-i k
\varphi} \sum_{n \in \mathbb{N}} \int_{-\infty}^\infty
\frac{d\omega}{\omega \Omega}\:e^{-i \omega t} \sum_{a,b=1}^2 t^{k
\omega n}_{ab}\: \Psi^a_{k \omega n}(r,\vartheta)\; \bra \Psi^b_{k
\omega n}, \Psi_0 \ket\:,
\end{equation}
where
\beq \label{o0def} 
\Omega(\omega) \;=\; \omega - \omega_0 \quad {\mbox{and}} \quad
\omega_0 \;=\; -\frac{ak}{r_1^2+a^2}\:.
\eeq
Here the sums and the integrals converge in~$L^2_{\mbox{\scriptsize{loc}}}$.
\end{theorem}
\noindent We now describe the various terms that appear in the
above integral representation in some detail, since they will be
used in our discussion of superradiance in Section
\ref{superradiance}. Recall from the discussion of the separation
of variables for the scalar wave equation that with the separation
ansatz \eqref{separansatz}, the scalar wave operator splits into
the sum of a radial operator \eqref{radial} and an angular
operator \eqref{angular}, with separated ordinary differential
equations \eqref{odes}. The angular
operator~${\mathcal{A}}_{\omega, k}$ has a purely discrete
spectrum of non-degenerate eigenvalues~$0 \leq \lambda_1 <
\lambda_2< \cdots$. The corresponding
eigenfunctions~$\Theta^{\omega,k}_n$ are the spheroidal wave
functions. In order to bring the radial equation into a convenient
form, we introduce a new radial function $\phi(r)$ by
\begin{equation}\label{rescal}
\phi(r) \;=\; \sqrt{r^2+a^2}\: R(r)\:, \end{equation}
and define the Regge-Wheeler variable~$u \in \mathbb{R}$ by
\begin{equation} \label{51a}
\frac{du}{dr} \;=\; \frac{r^2+a^2}{\Delta} \:,
\end{equation}
mapping the event horizon to $u=-\infty$. The radial equation then
takes the form of a Schr{\"o}dinger equation,
\begin{equation}
\left(-\frac{d^2}{d u^2} + V(u) \right) \phi(u) \;=\; 0\:.
\label{5ode}
\end{equation}
For fixed separation constants $k>0$, $\omega$ and $\lambda_n$,
the potential $V(u)$ has the asymptotics
\[ \lim_{u \to -\infty} V(u) \;=\; -\Omega^2 \:,\spc
\lim_{u \to \infty} V(u) \;=\; -\omega^2\:. \] Thus there are
fundamental solutions~$\acute{\phi}$ and~$\grave{\phi}$
of~\eqref{5ode} which behave asymptotically like plane waves
at the event horizon and near infinity, respectively,
\[ \acute{\phi}(u) \;\sim\; e^{i \Omega u} \;\;\;{\mbox{as $u \rightarrow -\infty$}},\qquad
\grave{\phi}(u) \;\sim\; e^{-i \omega u} \;\;\;{\mbox{as $u \rightarrow \infty$}} \]
(these functions are the so-called Jost solutions,
see~\cite{FKSY06} for details). As the corresponding time dependent solutions
behave in time like the plane wave~$e^{-i \omega t}$, these solutions
have a physical interpretation as incoming and outgoing waves in the asymptotic regions.
The functions~$\grave{\phi}$
and~$\overline{\grave{\phi}}$ form a fundamental system, and thus
we can represent~$\overline{\acute{\phi}}$ as
\begin{equation}\label{decompphi}
\overline{\acute{\phi}(u)} \;=\; A\: \grave{\phi}(u) + B \overline{\grave{\phi}}(u) \:.
\end{equation}
The coefficients~$A$ and~$B$ are the reflection and transmission coefficients.
The quantities~$t_{ab}$ in the integral representation \eqref{intrep} are explicit
functions of~$A$ and~$B$. Finally the functions $\Psi^a_{k \omega
n}(r,\vartheta),\,a=1,2$ are the solutions of the wave equation
\eqref{wave}, with fixed quantum numbers $k,\omega,n $,
corresponding to the real-valued fundamental solutions of the
radial equation given by \beq \label{phi12}
\phi^{\,1}={\mbox{Re}}\,{\acute{\phi}}\;,\qquad
\phi^{\,2}={\mbox{Im}}\, {\acute{\phi}}\:. \eeq

Our next result is a decay theorem for scalar waves in Kerr
geometry analogous to Theorem \ref{decay} for the Dirac operator.
\begin{theorem} \label{thmmain}
Consider the Cauchy problem for the wave equation in the Kerr
geometry for smooth initial data which is compactly supported
outside the event horizon and has fixed angular momentum in the
direction of the rotation axis of the black hole, i.e.\ for some~$k
\in \mathbb{Z}$,
\[ (\Phi_0, \partial_t \Phi_0) \:=\: e^{-i k \varphi} \:(\Phi_0, \partial_t \Phi_0)(r, \vartheta)
\;\in\; C^\infty_0((r_1, \infty) \times S^2)^2\:. \] Then the
solution decays in~$L^\infty_{\mbox{\scriptsize{loc}}}((r_1, \infty)
\times S^2)^2$ as~$t \to \infty$.
\end{theorem}
\noindent Theorem \ref{thmmain} can be interpreted as a
linear stability result for the Kerr black hole under
perturbations by massless scalar fields for a finite number of azimuthal angular momentum
modes. It would be desirable to analyze the convergence of the sum over all $k$-modes.
For this, one must obtain estimates which are uniform in~$k$. Such estimates have not
worked out, except in the case~$a \ll m$, a perturbation of the Schwarzschild space-time;
see~\cite{DR1}, where the authors obtain a boundedness result using the vector field method.
No theorems are known on rates of decay or probability estimates for scalar fields that
would be analogous to Theorems \ref{decayrates} and \ref{probest}
for Dirac fields. Proving such theorems in Kerr geometry appears
to be a significant challenge.

We conclude this section by sketching the proof of Theorem~\ref{intrepscal}. As for the
Dirac equation, we reformulate the wave equation in Hamiltonian
form. Letting
\begin{equation}
\Psi=\left(
\begin{array}{c} \Phi \\
i\partial_{t}\Phi
\end{array} \right),
\end{equation}
the wave equation \eqref{wave} takes the form
\begin{equation}\label{Hamform}
i\,\partial_{t}\Psi=H\,\Psi,
\end{equation}
where $H$ is the Hamiltonian
\begin{equation} \label{hamil}
H=\left(
\begin{array}{cc}
0 & 1 \\
\alpha & \beta
\end{array}
\right),
\end{equation}
and where $\alpha$ and $\beta$ are certain 
differential operators.  The general strategy for constructing an
integral representation for the propagator is the same as in the
case of the Dirac operator, and requires as a first step that the
problem be set up in a finite radial box, with Dirichlet boundary
conditions on the walls of the box. Unfortunately, the lack of a
positive conserved energy makes it impossible to associate a
positive definite invariant inner product to the time evolution of
the Hamiltonian, and one has to consider the Hamiltonian $H$ as
acting on a function space endowed with an inner product of {\em
indefinite} signature. In the case of a finite box, the
inner product space is a {\em Pontrjagin space} \cite{Bo}, that is
a complex vector space ${\mathcal{K}}$ endowed with a
non-degenerate inner product $\bra .\:,\: .\ket$ and an orthogonal
direct sum decomposition ${\mathcal{K}}={\mathcal{K}}_{+}\oplus {\mathcal{K}}_{-}$,
such that $({\mathcal{K}}_{+}, \bra .\:,\: .\ket)$ and
$({\mathcal{K}}_{-}, -\bra .\:,\: .\ket)$ are both Hilbert spaces,
with ${\mathcal{K}}_{-}$ being finite-dimensional. Classical
results of Pontrjagin \cite{Bo} imply that any self-adjoint
operator $A$ on a Pontrjagin space has a spectral decomposition,
which is similar to the one given by the spectral theorem in
Hilbert spaces, except that there is in general an additional
finite point spectrum in the complex plane (which is symmetric
about the real axis). We now consider the vector space
${\mathcal{P}}_{r_L,r_R}=(H^{1,2}\oplus L^2)([r_L,r_R]\times S^2)$
with Dirichlet boundary conditions $\Psi_1(r_L) = 0 = \Psi_1(r_R)$.
We endow this vector space with the inner product associated to the
energy \eqref{energy}. It can be shown \cite{FKSY05} that for every $r_R>r_1$ there is a
countable set $E\subset (r_1,r_R)$ such that for all $r_L \in
(r_1,r_R)\setminus E$, the inner product space
${\mathcal{P}}_{r_L,r_R}$ is a Pontrjagin space with the same
topology as $(H^{1,2}\oplus L^2)([r_L,r_R]\times S^2)$, and that
the Hamiltonian on the Pontrjagin space ${\mathcal{P}}_{r_L,
r_R}$ with domain ${\mathcal{D}}= C^\infty([r_L, r_R] \times
S^2)^2 \subset {\mathcal{P}}_{r_L, r_R}$ is essentially
self-adjoint. If the size of the box increases, the number of
complex conjugate pairs becomes larger, and in the infinite
volume limit all these spectral points move onto the real axis.
This is made precise using contour methods and estimates for the
resolvent. Whiting's mode stability result~\cite{W},
which guarantees the absence of unstable exponentially growing modes for the
separated wave equation, is crucial to our proof.

\section{A Rigorous Treatment of Superradiance for Scalar Waves}\label{superradiance}
One of the most fascinating aspects of the classical physics of
black holes is given by the so-called {\em Penrose process}
\cite{Penrose}, which shows that one can
extract energy and angular momentum from the Kerr black hole so as to
lower the angular momentum to zero by suitably exploiting the effect
of the ergosphere on the dynamics of point particles. The
basic idea of the Penrose process is as follows (see~\cite{Wald} for more
details and~\cite{Wagh} for more realistic scenarios involving collisions
of charged particles). First recall that the
conserved energy of a point particle of
momentum~$p$ is given by~$\langle p, \partial_t \rangle$,
and that this inner product clearly need not be positive inside the ergosphere,
where the vector field~$\partial_t$ is space-like.
Now consider a rocket which flies into the ergosphere, where
it splits into two objects whose energies have opposite signs.
By a suitable choice of the energy and momenta, one can arrange that
the object of negative energy crosses the event horizon and
reduces the energy and angular momentum of the black hole, whereas the other
object escapes to infinity, carrying (due to energy conservation)
more energy than the original rocket. In this way, one can extract
energy from the black hole, at the expense of reducing its angular
momentum. Christodoulou~\cite{Christo} showed that the infinitesimal
changes of mass $\delta M$ and angular momentum $\delta(aM)$ of the
black hole satisfy the inequalities \beq \label{deltaM} \delta (aM)
\;\leq\; \frac{r_1^2+a^2}{a} \:\delta M \;<\; 0\:, \eeq and as a
consequence he showed that it is not possible to reduce the mass of
the black hole via the Penrose process below the so-called {\em{irreducible
mass}}
\beq \label{Mirr}
M_{\mbox{\scriptsize{irr}}}^2 \;:=\; \frac{1}{2} \left( M^2 +
\sqrt{M^4 -(aM)^2} \right)\:. \eeq

\begin{figure}[tbp]
\begin{center}
\begin{picture}(0,0)%
\includegraphics{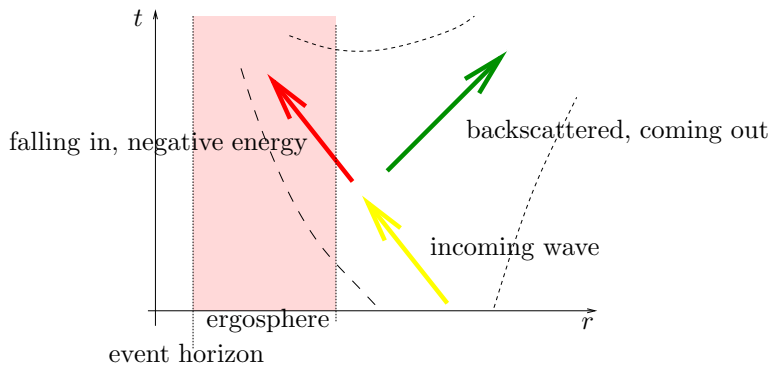}%
\end{picture}%
\setlength{\unitlength}{829sp}%
\begingroup\makeatletter\ifx\SetFigFont\undefined%
\gdef\SetFigFont#1#2#3#4#5{%
  \reset@font\fontsize{#1}{#2pt}%
  \fontfamily{#3}\fontseries{#4}\fontshape{#5}%
  \selectfont}%
\fi\endgroup%
\begin{picture}(16582,10606)(-2699,-9745)
\put( 16,359){$t$}
\put(-721,-9691){\normalsize{event horizon}}
\put(2206,-8656){\normalsize{ergosphere}}
\put(13411,-8701){$r$}
\put(-3699,-3346){\normalsize{falling in, negative energy}}
\put(8886,-6496){\normalsize{incoming wave}}
\put(9966,-2986){\normalsize{backscattered, coming out}}
\end{picture}%
\caption{Superradiant scattering.}
\label{figsuper}
\end{center}
\end{figure}

Superradiance is the wave analogue of the Penrose process.
One considers a wave entering the black hole. One part of the wave enters
the event horizon, whereas the other part is scattered at the black
hole and gives rise to an outgoing wave (see Figure~\ref{figsuper}).
By arranging the energy of the infalling wave to be negative, one can
again extract energy from the black hole.
This effect was first studied by
Zel'dovich~\cite{Z} and Starobinsky~\cite{S} on the level of modes, i.e.\
by considering the transmission and reflection coefficients in~\eqref{decompphi}
associated to the Schr\"odinger equation~\eqref{5ode}. In this analysis,
the quantities~$\omega^2 |A|^2$ and~$\omega^2 |B|^2$ have
the interpretations as the energy flux of the incoming and outgoing
waves, respectively. Thus the relative energy gain~${\mathfrak{R}}$ is given by
\beq \label{frakRdef} {\mathfrak{R}} \;=\; \frac{|B|^2}{|A|^2} \:.
\eeq
Computing the Wronskians of~$\phi$ and
$\overline{\phi}$ near the event horizon and near infinity gives the
relation
\beq \label{ABeq} |A|^2 - |B|^2 \;=\; \frac{\Omega}{\omega} \:.
\eeq
If the right side of~\eqref{ABeq} is positive,
the outgoing flux is smaller than the incoming flux, and this
corresponds to ordinary scattering. However, if the right side
of~\eqref{ABeq} is negative, then the outgoing flux is larger than
the incoming flux, and according to~\eqref{frakRdef} we gain energy.
This is termed {\em{superradiant scattering}}.
Using~\eqref{o0def}, superradiant scattering appears precisely
when $\omega$ is in the range
\beq \label{omegasuper}
0 \:<\; |\omega| \;<\; |\omega_0|\:.
\eeq
Starobinsky~\cite{S} computed~${\mathfrak{R}}$ and found a relative gain of energy of
about~$5\%$ for $k=1$ and less than~$1\%$ for $k \geq2$. Teukolsky and
Press~\cite{TeP} made a similar mode analysis for higher spin and
found numerically an energy gain of at most~$4.4\%$ for Maxwell
($s=1$) and up to $138\%$ for gravitational waves ($s=2$).
For more recent developments on the mode analysis of superradiance and
related physical effects we refer to~\cite{Cardoso}.

Unfortunately, the mode analysis does not give information on the dynamics.
Thus for a rigorous treatment of energy extraction one needs to consider the
time-dependent situation. This was done numerically in~\cite{ALP}
for wave packet initial data. 
The main result of \cite{FKSY07} treats the time-dependent situation
rigorously for the Cauchy problem. As initial data we take wave packets
of the form
\begin{eqnarray}\label{wavepack}
\Psi_{0} \;=\; \Theta_{{\tilde n},{\tilde \omega}}(\vartheta)\:e^{-i k \varphi}\:
\frac{\eta_{L}(u)}{\sqrt{r^{2}+a^{2}}} \left[ c_\text{in}\,e^{-i{\tilde
\omega}u}  \left( \!\begin{array}{cc} 1 \\ {\tilde \omega}
\end{array}\! \right) + c_\text{out}\,e^{i{\tilde \omega}u}  \left(
\!\begin{array}{cc} 1 \\ -{\tilde \omega} \end{array}\! \right)
\right] \:,
\end{eqnarray}
where~$L$ is a large parameter, and~$\eta_L$ is a smooth cutoff function
of the form
\[ \eta_{L}(u)\;=\;\frac{1}{\sqrt L}\,\eta \!\left(\frac{u-L^2}{L} \right) , \]
with $\eta \in C^\infty_{0}(\mathbb{R}_{+})$. Here~$\Theta_{{\tilde n},{\tilde \omega}}(\vartheta)$ is an eigenfunction of the angular operator ${\mathcal A}$.
The energy radiated to infinity is defined by
\beq \label{Eoutdef}
\Eout \;=\; \lim_{t \to \infty} \bra \Psi(t), \chi_{(2r_1,
\infty)}(r)\, \Psi(t)\ket \:,
\eeq
where $\chi$ is the characteristic function.

\begin{theorem}\label{maintheo} For any~$R>r_1$
and~$\delta>0$ there is initial data $\Psi_0 \in C^\infty_0((R,
\infty) \times S^2)^2$ of the form~\eqref{wavepack}
such that the limit in~\eqref{Eoutdef} exists and
\[ \left| \frac{\Eout}{\bra \Psi_0, \Psi_0 \ket}
- {\mathfrak{R}} \right| \;\leq\; \delta \] with~$\mathfrak{R}$ as
in~\eqref{frakRdef}.
\end{theorem}
\noindent
For the proof we consider the integral representation of Theorem~\ref{intrepscal}
for the wave packet initial data~\eqref{wavepack}. 
The crucial analytical ingredient in the proof
is the time-independent energy estimate for the outgoing wave as derived in~\cite{FS}. 

It should be stressed that we allow the initial data to be supported
arbitrarily far away from the event horizon. This is important in
order to avoid artificial initial data which would not correspond to
an energy extraction mechanism. For example, if one allows the
support of the initial data to intersect the ergosphere, one could
take initial data with zero total energy, in which case the
quotient~$\Eout/\bra \Psi_0, \Psi_0 \ket$ could be made arbitrarily
large. 

\section{The Stability Problem for Kerr Black Holes}
The remaining challenge is to prove the {\em linear
stability} of the Kerr black hole under electromagnetic and gravitational 
perturbations. As Frolov and Novikov put it~\cite[S.~143]{FN},
this is ``one of the few truly outstanding problems that remain in the field of black
hole perturbations.''
 Since the linear perturbations of the Kerr metric
are described by the solutions of the Teukolsky equation
\eqref{Teukolsky}, proving linear stability amounts to showing
that the solutions of the Cauchy problem for the Teukolsky
equation for compactly supported Cauchy data decay in
$L^{\infty}_{loc}$ as $t$ tends to infinity. In other words, the
task is to prove the analogue of Theorem~\ref{thmmain} for
the solutions of the Teukolsky equation~\eqref{Teukolsky}
in the cases~$s=1$ and~$s=2$.
The analysis of the Teukolsky equation for higher spin appears to be considerably more
difficult than that of the scalar wave equation.
This is because, in contrast to the scalar wave equation~\eqref{Lagrange},
the Teukolsky equation has no variational formulation, and thus there is no
simple method for deriving the conserved quantities.
Indeed, the expressions for the physical energy and charge are very complicated
and involve higher derivatives of the field~$\Psi_s$.
The only result so far on the Teukolsky equation in Kerr is the deep and important
mode stability theorem proved by Whiting in~\cite{W}.

More recently, a rigorous stability result was proved for the Teukolsky equation
in the simpler setting of the Schwarzschild geometry~\cite{FS0} for electromagnetic
and gravitational wave perturbations, which we now describe.
The evolution of a massless wave of spin~$s$ in the Schwarzschild geometry is described
by the Teukolsky equation (set $a=0$ in~\eqref{wave})
\begin{eqnarray}
\lefteqn{ \left[ \partial_r \Delta \partial_r - \frac{1}{\Delta} \left( r^2 \partial_t - (r-M) s \right)^2
- 4 s r \partial_t \right. } \nonumber \\
&& \left. + \partial_{\cos \vartheta} \sin^2 \vartheta\:
\partial_{\cos \vartheta}
+ \frac{1}{\sin^2 \vartheta} \left(\partial_\varphi + i s \cos \vartheta \right)^2 \right] \Phi(t,r,\vartheta, \varphi) \;=\; 0 \:, \label{teq}
\end{eqnarray}
where~$\Delta = r^2-2Mr$. We consider~\eqref{teq} with~$C^\infty_0$ initial data
\beq \label{72}
\Phi |_{t=0} \;=\; \Phi_0\:,\quad \partial_t \Phi|_{t=0} \;=\; \Phi_1\:.
\eeq
Then the following theorem holds (see~\cite{FS0}).
\begin{theorem} For spin~$s=1$ or~$s=2$, the solution of the Cauchy problem~\eqref{teq},
\eqref{72} for~$(\Phi_0, \Phi_1) \in C^\infty_0((2M, \infty) \times S^2)^2$ decays
in~$L^\infty_{\mbox{\scriptsize{loc}}}((2M, \infty) \times S^2)$ as~$t \rightarrow \infty$.
\end{theorem}
The proof has some novel features, and we shall discuss some of them.

After separating the time and angular dependence similar to~\eqref{separansatz}, \eqref{odes}
(with~$k \in \Z$, $\lambda \in \mathbb{R}$ and $\omega \in \C$)
and replacing~$r$ by the Regge-Wheeler variable~$u$, the Teukolsky equation can be
reduced to a one-dimensional Schr\"odinger equation
\[ -\frac{d^2}{du^2} \,\phi(u) + V(u)\, \phi(u) \;=\; 0 \]
with potential~$V$ given by
\[ V(u) \;=\; -\omega^2 + is \omega \left[ \frac{2 (r-M)}{r^2} - \frac{4 \Delta}{r^3}
\right] + \frac{(r-M)^2\, s^2}{r^4} + \frac{\partial_u^2 r}{r} + \lambda\: \frac{\Delta}{r^4}\:, \]
where~$\lambda$ is an eigenvalue of the angular operator. Due to the presence of the
$\partial_t$-term in~\eqref{teq}, $V$ is complex even for real~$\omega$. Thus most standard
techniques are unavailable.

\begin{figure}[tbp]
\begin{center}
\begin{picture}(0,0)%
\includegraphics{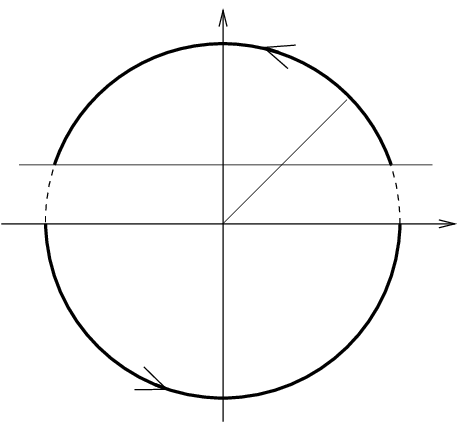}%
\end{picture}%
\setlength{\unitlength}{1243sp}%
\begingroup\makeatletter\ifx\SetFigFont\undefined%
\gdef\SetFigFont#1#2#3#4#5{%
  \reset@font\fontsize{#1}{#2pt}%
  \fontfamily{#3}\fontseries{#4}\fontshape{#5}%
  \selectfont}%
\fi\endgroup%
\begin{picture}(7102,6510)(2904,-6385)
\put(9226,-2806){\makebox(0,0)[lb]{\smash{\SetFigFont{12}{14.4}{rm}$S_2$}}}
\put(3016,-2806){\makebox(0,0)[lb]{\smash{\SetFigFont{12}{14.4}{rm}$S_1$}}}
\put(7381,-6271){\makebox(0,0)[lb]{\smash{\SetFigFont{12}{14.4}{rm}$C_1$}}}
\put(7246,-2761){\makebox(0,0)[lb]{\smash{\SetFigFont{12}{14.4}{rm}$R$}}}
\put(9991,-2401){\makebox(0,0)[lb]{\smash{\SetFigFont{12}{14.4}{rm}$\displaystyle {\mbox{Im}}\, \omega = \frac{s}{4M}$}}}
\put(7336,-196){\makebox(0,0)[lb]{\smash{\SetFigFont{12}{14.4}{rm}$C_2$}}}
\end{picture}%
\caption{The contour~$C_R$.}
\label{figcontour}
\end{center}
\end{figure}

First one easily checks that
\[ \lim_{u \rightarrow -\infty} V(u) \;=\; -\left(\omega - \frac{is}{4M} \right)^2\:. \]
Then writing the time-dependent equations in Hamiltonian form,
\[ i \partial_t \Psi \;=\; H\, \Psi \:,\qquad \Psi = (\Phi, \partial_t \Phi)\:, \]
with~$H$ a matrix differential operator, we can show that if~$\omega$ is outside
the strip $0 \leq {\mbox{Im}}\, \omega \leq \frac{s}{4M}$, then~$\omega$ lies in the
resolvent set of~$H$, and the resolvent~$R_\omega$ is holomorphic there.
By means of suitable resolvent estimates, we show that any~$\Psi \in C^\infty_0$
has the integral representation
\[ \Psi(u) \;=\; -\frac{1}{2 \pi i} \lim_{R \rightarrow \infty}
\int_{C_R} (R_\omega \Psi)(u)\: d\omega\:, \]
where~$C_R$ is the contour shown in Figure~\ref{figcontour}, which has
two connected components~$C_1$ and~$C_2$. 
Using contour deformation techniques, we can prove the following non-standard
integral representation theorem for the solution.

\begin{theorem}
For spin $s=1$ or~$s=2$,
the solution of the Cauchy problem for the Teukolsky equation~\eqref{teq}
with initial data~$\Psi_0 = (\Phi, \partial_t \Phi)|_{t=0} \in C^\infty_0(\R)^2$
has the representation
\begin{eqnarray*}
\Psi(t,u) &=& -\frac{1}{2 \pi i} \int_{\R} e^{i \omega t}
\left( ({\mathfrak{R}}_\omega \Psi_0)(u) + \frac{\Psi_0(u)}{\omega+i} \right) d\omega \\
&&+\frac{1}{2 \pi i} \int_{\R + \frac{i s}{2M}}  e^{i \omega t} \left(
(R_\omega \Psi_0)(u) + \frac{\Psi_0(u)}{\omega+i} \right) d\omega\:,
\end{eqnarray*}
where~$\mathfrak{R}_\omega$ is the limit of the integral kernel of the resolvemt
from the lower half plane,
\[ {\mathfrak{R}}_{\omega_0}(u,v) \;:=\;
\lim_{\omega \rightarrow \omega_0,\; {\mbox{\rm{\scriptsize{Im}}}}\, \omega < 0}
R_\omega(u,v)\:. \]
Both integrands in the above integral representation are in $L^1$.
\end{theorem}
\noindent
Decay is an immediate consequence of this theorem: For~$t \rightarrow \infty$, the first
integral tends to zero by the Riemann-Lebesgue lemma, while the second integral tends
exponentially to zero.

\bibliographystyle{amsalpha}

\end{document}